\documentclass{epl}
\usepackage{graphicx}
\usepackage{amssymb}
\usepackage{amsmath}

\title{Elastic manifolds in disordered environments: energy statistics}

\author{K. P. J. Kyt\"ol\"a\inst{1} \and E. T. Sepp\"al\"a\inst{2} \and
M. J. Alava\inst{1}}
\institute{
  \inst{1} Helsinki University of Technology, Laboratory of Physics,
P.O.Box 1100, FIN-02015 HUT, Finland \\ 
  \inst{2} Lawrence Livermore National Laboratory, 7000 East Avenue, 
L-415, Livermore, CA 94550, U.S.A.
}


\begin{document}

\maketitle

\begin{abstract}
The energy of an elastic manifold in a random landscape at $T=0$ is
shown numerically to obey a probability distribution that depends on
size of the box it is put into. If the extent of the spatial
fluctuations of the manifold is much less than that of the system, a
cross-over takes place to the Gumbel-distribution of extreme
statistics.  If they are comparable, the distributions have
non-Gaussian, stretched exponential tails.  The low-energy and
high-energy stretching exponents are roughly independent of the
internal dimension and the fluctuation degrees of freedom.
\end{abstract}

The statistical mechanics of elastic objects or manifolds changes in
the presence of disorder, since temperature becomes irrelevant as a
scaling variable. The statistical properties are determined by the
competition of elasticity and randomness.  Examples of systems where
this happens are domain walls (DW) in random magnets and flux lines in
superconductors~\cite{blatter94,Hah95,Fis86}.  At the zero temperature
``fixed point'' in the renormalization group language, the geometry of
the object becomes critical.  The fluctuations are self-affine below
the upper critical dimension of the Hamiltonian.  This upper critical
dimension may or may not have a finite value depending on the
dimensionality of the possible fluctuations.  Examples are provided by
the one-dimensional directed polymer (DP) in $n \geq 1$ fluctuation
dimensions, which has a low-temperature phase for which $n_c =\infty$
\cite{Hah95}.  Another case is given by $D$ internal-dimensional
random manifolds in $d=(D+n)$ dimensions, where $n=1$, which are known
to have the upper critical dimension $D_c=4$~\cite{Fis86}.

The criticality of elastic manifolds is manifested by the distribution
of energies, $P(E)$, which is not a Gaussian. This was shown by
extensive simulations of directed polymers in $d=(D+n)= (1+1)$ and
(1+2) dimensions ($d$ is the total, embedding space dimension of a
system)~\cite{KBM}.  In these simulations a DP was let to minimize its
energy by keeping one end fixed, and letting the other one wander
freely. The outcome is that for energies smaller than the average,
$E_- \ll \langle E \rangle$, the distribution $P(E) \sim
\exp[-|E|^{\eta_-}]$, {\it i.e.}, {\em stretched exponential}. Note
that the distribution is normalized in such a way that the average of
$E$ is zero and its variance or standard deviation equals
unity. Similarly for $E_+ \gg \langle E \rangle$ one obtains a
stretching exponent $\eta_+$. These stretching exponents are below the
Gaussian $\eta= 2$ value for small energies and above it for large
ones, and their numerical values are rather insensitive to whether
$n=1$ or 2.

What has not been addressed so far is how
exactly the scaling function, $P(E)$, depends on the set of boundary
conditions imposed on the manifold and the dimensions $n$ and $D$. One
reason is that computing $P(E)$ is naturally a rather formidable
task. One rare analogy is provided by the (1+1)-dimensional
Kardar-Parisi-Zhang-equation (KPZ) \cite{KPZ}, in the form of the
asymmetric exclusion process (ASEP).  This system is governed by the
``strong coupling fixed point'' of the KPZ interface
growth. Approaches based on a generalized free energy by Derrida and
Lebowitz~\cite{Derr98} and on mapping the problem to random matrix
theory (by Pr\"ahofer and Spohn~\cite{spohn}) have succeeded in
deriving analytically the non-Gaussian scaling functions. This is true
for the velocity fluctuations or height fluctuations in the two
respective cases but only in the {\em steady-state} of the growth,
that is when the correlation length exactly equals the system size.
In the KPZ case in (1+1)-dimensions one can also resort to
scaling arguments for the extremal properties of the interface
statistics, predicting $\eta_- = 3/2$ \cite{redner}.

In this Letter we show that the problem of the shape of the
probability distribution $P(E)$ is actually continuously dependent on
a natural boundary condition, namely the transverse height of the
system.  It is also, on the other hand, rather independent of the
dimensionality.  The continuum of exponents $\eta_+$, $\eta_-$ for
$P(E)$ results from the cross-over from a Gaussian distribution (when
the height of the system is limited to below the natural roughness or
geometrical fluctuations and hence reflects Poissonian statistics) to
the Gumbel function of extreme statistics~\cite{Galambos,boumez} when
the transverse height of the system is let to increase to
infinity. It turns out to be so that the variation of the $\eta_-$,
$\eta_+$ is related to so-called penultimate extremal distributions,
the scaling of an extremal quantity from samples with a {\em finite}
number of independent variables. 

There remain several open questions that we return to in the
summary.  Recall that the energy of a DP relates to the first arrival
time of a KPZ interface to a fixed height (a related problem but not
quite the same is the maximal height of an interface with a given
average height \cite{sha01}).  Hence varying the
transverse height thus equals moving between the stationary state and
an initial transient. The former can be accessed via the large
deviation and random matrix formalisms \cite{Derr98,spohn}.  In our
numerical simulations we have varied both the internal dimension of
the system $D$, which gets values from one to three, and the
fluctuation degree of freedom $n$, which gets the same values. We are
for algorithmic reasons restricted to cases, where whenever $D>1$ $n$
has to have a value of unity, and vice versa.

The geometric fluctuations of elastic manifolds can be measured by the
two-point correlation function or by mean-square fluctuations. These
define the roughness exponent $\zeta$, {\it e.g.} as $w^2 = \langle
[z({\bf x}) - \overline{z({\bf x})} ]^2 \rangle \sim L^{2 \zeta}$,
where $L$ is the linear size of the system, $z({\bf x})$ is the
manifold's height and ${\bf x}$ is the $D$ dimensional internal
coordinate of the manifold. For the ground state or free energy
fluctuations around the disorder-averaged mean one has that $\Delta E
= \left \langle ( E - \langle E \rangle )^2 \right \rangle^{1/2} \sim
L^\theta$, where $\theta = 2 \zeta +D -2$~\cite{HuHe}. The ground
state energy is a quenched random variable and it behaves as $E \sim
AL^D + B L^\theta$.  Thus normal equilibrium behavior is modified, and
the (free) energy develops non-analytic corrections. In a
renormalization picture the energy fluctuation exponent $\theta$
arises naturally.  We concentrate on random-bond disorder, which means
that the random-potential in the energy Hamiltonian is delta-point
correlated, although the results should be applicable to other types
of short-range correlated randomness, too. At low temperatures in
$(1+1)$ dimensions, due to the equivalence of the directed polymer in
random media~\cite{KPZ,Hah95} to the KPZ equation, the exact roughness
exponent is $\zeta=2/3$, with RB disorder~\cite{KPZ,Hah95,Barabasi}.
In higher $D$ dimensions functional renormalization group calculations
give $\zeta \simeq 0.208(4-D)$~\cite{Fis86}. The $(1+n)$ dimensional
directed polymers with $n>1$ have roughness values $\zeta=0.62$ and
$\zeta=0.59$, when $n=2$ and $n=3$, respectively (see {\it e.g.}
\cite{Hah95}).

\begin{figure}
\onefigure[width=6cm]{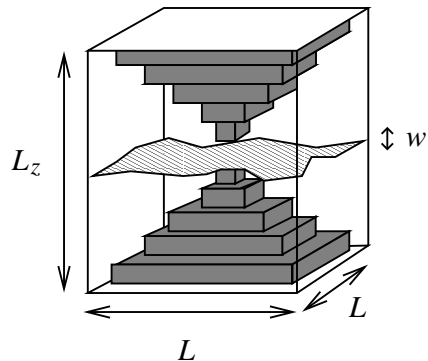}
\caption{A sketch of $(D+n) =(2+1)$ dimensional manifold in a system,
whose lateral size, perpendicular to the average normal of the
surface, is $L_x = L_y =L$ and transverse height, parallel to the
average surface normal, is $L_z$. The width, root-mean-square
fluctuation, of the surface is $w$. When the manifold is pinned in the
center of the system, the ``cone'' geometry is used. In that case the
area where the surface is not allowed to pass through is depicted with gray.}
\label{fig1}
\end{figure}
For numerical calculations, when $D \geq 1$, the optimization of the
Hamiltonian of a manifold in random bond environment (which is
equivalent with domain wall in random-bond Ising magnet) maps into the
minimum cut-maximum flow problem of combinatorial
optimization~\cite{Goltar88,Alavaetal}. The simulations of $D=1$,
$n\geq 1$ directed-polymers are done using the standard
transfer-matrix (TM) method~\cite{kardar87}. Here we impose two
different type of boundary conditions in the transverse height
direction of the system, {\it i.e.}, in the direction of the average
normal of the $D$ dimensional manifold. The manifold is fixed in a
single valley by pinning it at certain coordinate, or let to float
freely, so that the mean height can vary in a box defined by the
system height $L_z$.  For DP's the pinning follows in the TM
simulations from having one end fixed and the other one free. For
$D>1$ manifolds we construct a new way to create a ``one valley'' case
since the same trick as for DP's to fix the manifold at a fixed height
does not work, for dimensional reasons. Our pinning method is
illustrated in the (2+1)-dimensional case in Fig.~\ref{fig1} and is
also in the simulations extended to (3+1)-dimensions.

We have shown earlier that the average manifold energy develops a
logarithmic dependence on $L_z$~\cite{SAD01,Seppala01} due to extremal
statistics.  Since the width of a directed manifold grows as $L^\zeta$
it is expected that there is a number $N_z$ of quasi-independent
valleys in the energy landscape~\cite{Mez,Hwa,Seppala00}, proportional
to $N_z \sim L_z/L^\zeta$.  $N_z$ thus presents a scaling parameter,
and the Gumbel limit is obtained for $L$ fixed in the limit $L_z
\rightarrow \infty$, $N_z \rightarrow \infty$. Hence the logarithmic
dependence on $L_z$ results straightforwardly from extreme statistics.
We first study $P(E)$ in the case of the ``one valley'' systems, $N_z
=1$, and then in the case when $L_z$ or $N_z$ is varied.

\begin{figure}
\onefigure[width=50mm,angle=-90]{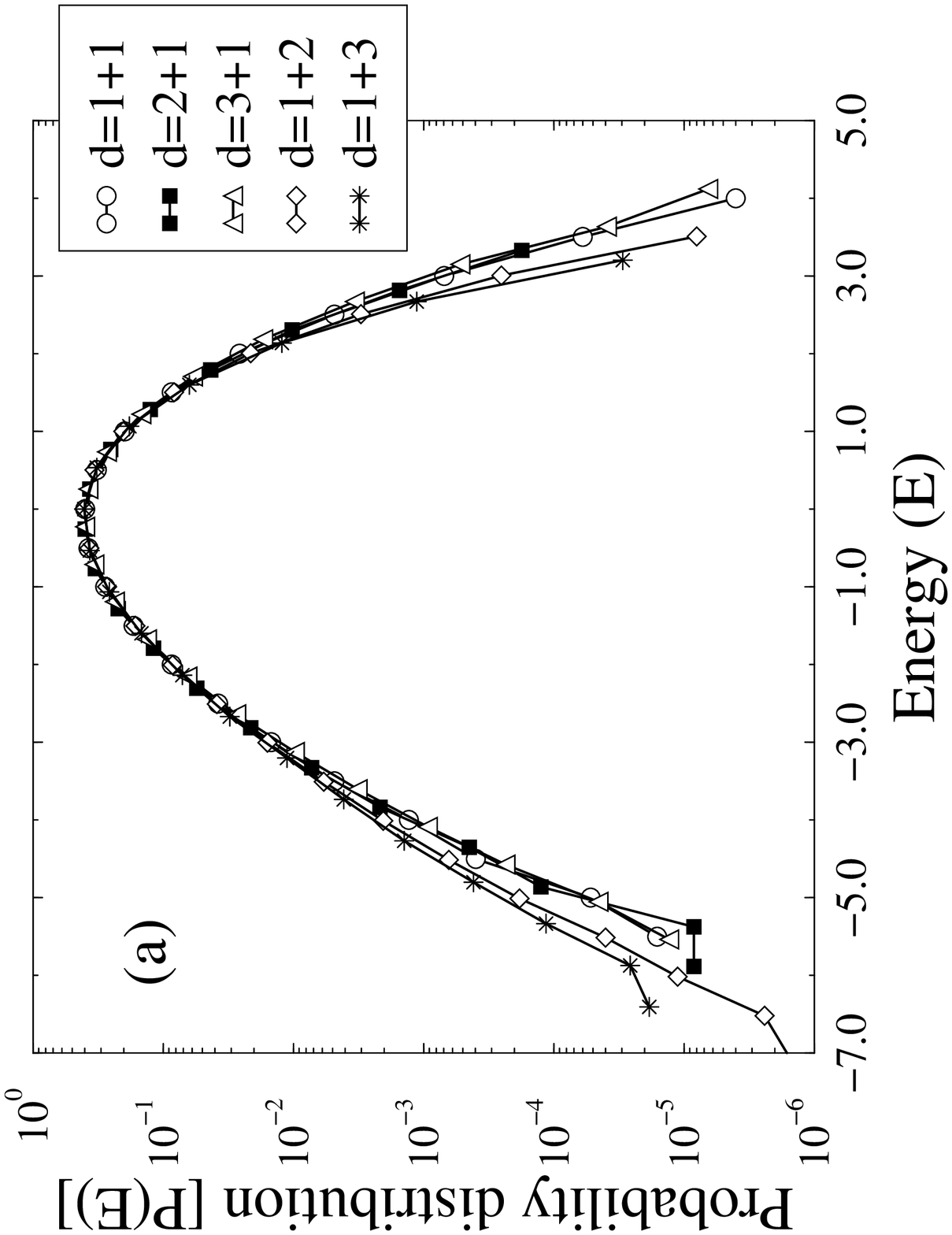}
\onefigure[width=50mm,angle=-90]{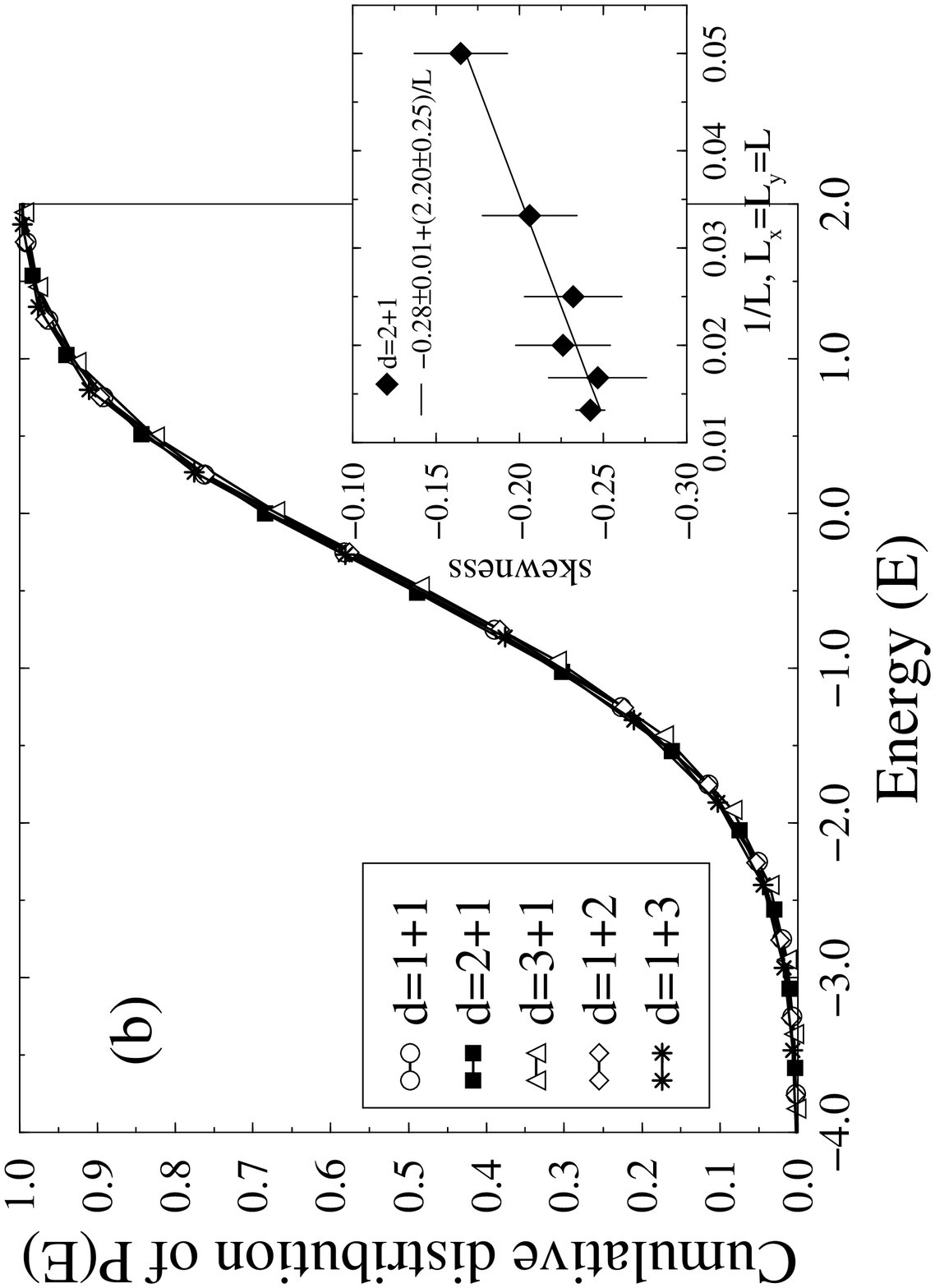}
\caption{(a) The normalized energy distributions, $\langle E
\rangle=0$ and $\Delta E =1$, for manifolds with various $D$ and $n$
in a ``single valley'' case, {\it i.e.}, pinned when $D=1$ or using
``cone'' geometry when $D>1$ ($L_z = L$).  The number of random configurations
$N=5\times 10^5$ and the lateral size of the system $L=100$ for $(D+n)
=(1+1)$. Similarly $N=2.3\times 10^5$ and $L^2 =75^2$ for $(D+n)
=(2+1)$; $N=3.2\times 10^5$ and $L^3 = 25^3$ for $(D+n) =(3+1)$;
$N=2.5\times 10^6$ and $L=70$ for $(D+n) =(1+2)$; $N=5.1\times 10^5$
and $L=50$ for $(D+n) =(1+3)$. (b) Cumulative distribution $\int P(E)
\, dE$ for the same data as in (a). The inset shows skewness, $\langle
[(E- \langle E \rangle) / \Delta E]^3 \rangle$, versus lateral size,
$L_x = L_y =L$, for $(D+n) = (2+1)$. The ``cone'' geometry is used in
the transverse direction. The solid line is the least-squares-fit to
the data leading to a value of $-0.28\pm 0.01$. The number of random
configurations $N=10^4$ for all the other $L$ except $L=75$ for which
the data is the same as in (a) and (b).}
\label{fig2}
\end{figure}
Figs.~\ref{fig2}(a) and \ref{fig2}(b) demonstrate the phenomenology of
$P(E)$ for various dimensionalities in the ``one valley'' case. We see
that the scaled probability distributions are rather independent of
$D$ and $n$ though the actual scaling exponents $\zeta$ and $\theta$
vary greatly. This is most striking if one considers the integrated
distribution [Fig.~\ref{fig2}(b)]. However, there we concentrate in
the part of the distribution, which is close to mean, and as the
actual shape is usually close to Gaussian the differences in the tails
cannot even be visible. One of the features of the behavior of $P(E)$
is that while the skewness, {\it i.e.}, 3rd moment, $\sigma_3 = \langle
[(E- \langle E \rangle) / \Delta E]^3 \rangle$, is $L$-dependent the
tail exponents are much less sensitive to the system size. The inset
demonstrates an extrapolation of skewness for (2+1)-dimensions, and we
obtain $\sigma_3 = -0.28 \pm 0.01$.

\begin{figure}
\onefigure[width=50mm,angle=-90]{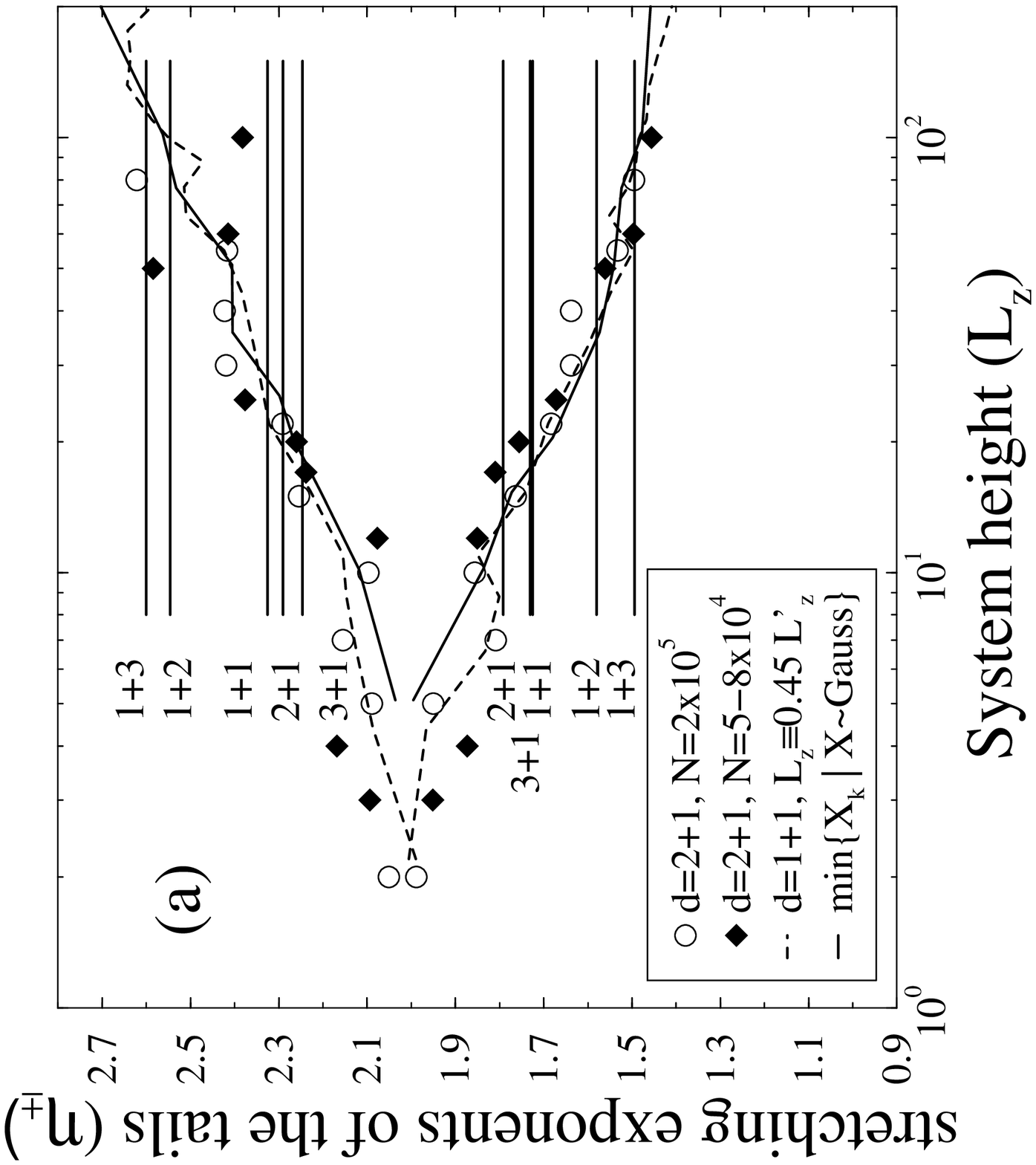}
\onefigure[width=50mm,angle=-90]{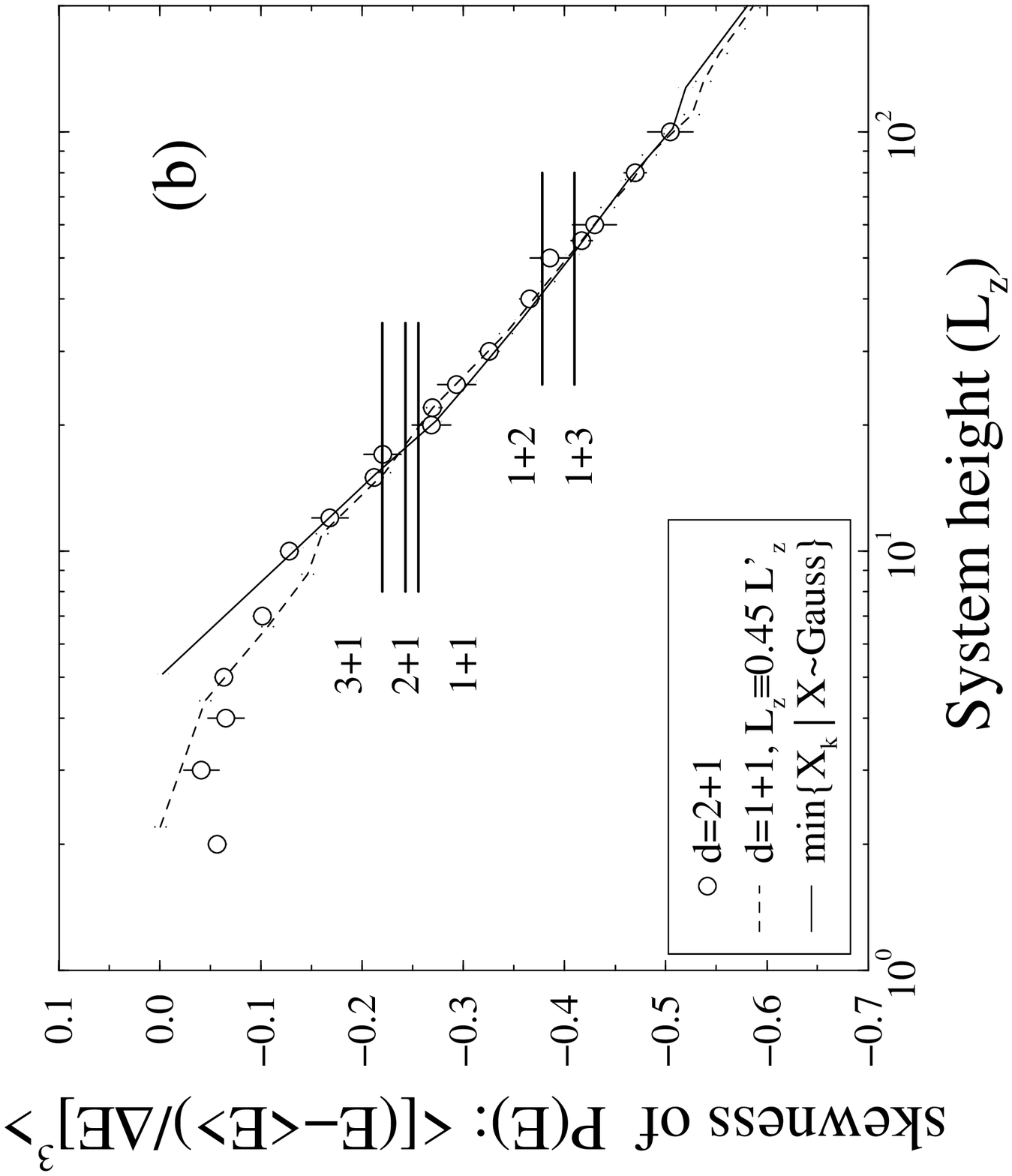}
\onefigure[width=50mm,angle=-90]{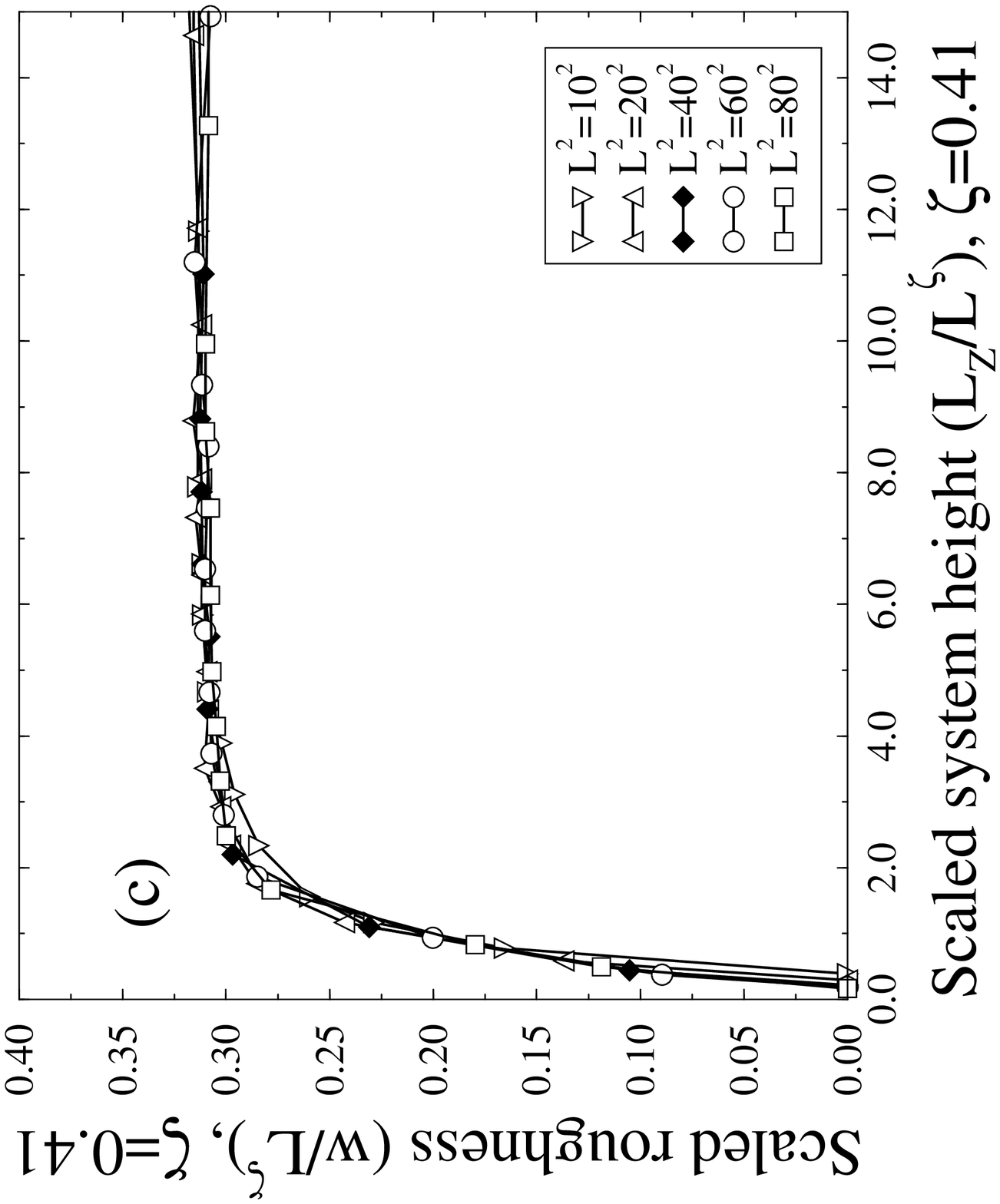}
\caption{(a) Stretching exponents, $\eta_-$ and $\eta_+$, of the left
and right-hand tails, respectively, of the energy distributions, $\exp
\{ -|(E- \langle E \rangle)/\Delta E|^{\eta_\pm} \}$, for $(D+n)
=(1+1)$ and $(2+1)$ versus the system height $L_z$ in unpinned (and
non-cone-geometry) cases. Open circles and filled diamonds are from
$(2+1)$ dimensional data. The first one has $N=2\times 10^5$, while
the latter has $N=5-8\times 10^4$. For both data set $L_x= L_y =
L=40$. $(D+n) =(1+1)$ is plotted as a dashed line. In that case
$L=100$. In order to show the similarity of the trends with $(D+n)
=(2+1)$ $L_z$ for $(1+1)$ are scaled by a factor of 0.45. The solid
line shows the stretching exponent for the distribution of the minimum
of several Gaussian distributions $\min\{X_k | X \sim Gauss,
k=1,2,\ldots, N_G\}$ versus the number of Gaussian distributions
$N_G$. In that case the $x$-axis is scaled so that $N_G = 5.1
L_z$. The horizontal lines show the stretching exponent values
$\eta_\pm$ of the distributions plotted in Fig.~\ref{fig2}(a). (b) The
skewness values of the same distributions as in (a). For $(D+n) =
(2+1)$ only the data with better statistics $N=2\times 10^5$ is
shown. For $(D+n) =(1+1)$ and for the Gaussian case we use the same
$x$-axis scalings as in (a).  The horizontal lines show the skewness
values of the distributions plotted in Fig.~\ref{fig2}(a). (c) The
scaled roughness $w/L^\zeta$ vs. scaled system height $L_z/L^\zeta$,
where $\zeta=0.41$, for $(D+n) =(2+1)$ and various lateral lengths of
the system $L_x= L_y = L =10$, 20, 40, 60, and 80. The number of realizations
varies from $N=1000$ for $L=10$ to $N=600$ for $L=80$.}
\label{fig3}
\end{figure}
Let us now consider what happens as a function of $N_z =L_z/w$.  If
$N_z \ll 1$ the interface is confined to a ``box'' which is smaller
than the expected roughness, thus criticality is destroyed and $P(E)$
becomes a Gaussian.  If $N_z$ is increased (by $L_z$) the outcome is
depicted in Fig.~\ref{fig3}(a). For all the varying dimensionalities
we obtain a steady but very slow cross-over of $P(E)$ towards a fast
decay on the side of large energies ($E \gg \langle E\rangle$) and a
slower one on the other side. In the asymptotic limit one expects that
the Gumbel distribution is found, which has the form (in scaled units)
$P_{Gumbel} \sim \exp{(E - \exp{E})}$.  The convergence is an example
of penultimate extremal distributions. For $N_z$ finite the extremal
probability-density-function (PDF) is given by a form which has a
point-wise limit in the asymptotic distribution \cite{books,refe},
but may e.g.~exhibit stretched exponential tails, such that the
behavior of the exponents is in agreement with the convergence
to the asymptotic. Thus $\eta_-$ increases, and $\eta_+$ decreases.
Penultimate distributions depend on many factors and can resemble
more the other two asymptotic forms (than the Gumbel) \cite{refe},
and the convergence can be very slow.

Next, one can pose the question: is the one-valley ensemble (with
appropriate boundary conditions) equivalent to what one gets in the
limit $N_z \simeq 1$, {\it i.e.}, are the $P$'s the same? This is not a
trivial one since the $N_z$-ensemble is ``grand-canonical'', the
number of ``valleys'' is not defined in any sample except in the
average sense (recall also the KPZ-growth analogy for DP's). Our
numerical results imply that this is roughly the case. We also
show that the convergence of the exponents is reproduced [as depicted in
Fig.~\ref{fig3}(a)] by a simple Gaussian model. We mimick a
single-valley distribution by picking the minimum out of a small number of
independent-and-identically-distributed Gaussian variables. Then $N_z$
is varied and the smallest of $N_z$ such ones is chosen. This gives
rise to the last set of data in the figure, and as can be seen it
reproduces the manifold results qualitatively and almost
quantitatively. This analogy reinforces the picture of independent
valleys and $N_z$ as a scaling parameter.

\begin{table}
\caption{The skewness $\sigma_3$ and stretching exponent values
$\eta_-$ and $\eta_+$ of the distributions plotted in
Fig.~\ref{fig2}(a).}
\label{tab}
\begin{center}
\begin{tabular}{l c c c c}
$D$ & $n$ & $\sigma_3$ & $\eta_-$ & $\eta_+$  \\
3 & 1 & $-0.22\pm0.01$ & $1.73\pm0.10$ & $2.25\pm0.10$ \\
2 & 1 & $-0.24\pm0.01$ & $1.79\pm0.10$ & $2.29\pm0.10$ \\
1 & 1 & $-0.26\pm0.01$ & $1.72\pm0.10$ & $2.33\pm0.10$ \\
1 & 2 & $-0.38\pm0.01$ & $1.58\pm0.10$ & $2.55\pm0.10$ \\
1 & 3 & $-0.41\pm0.01$ & $1.49\pm0.10$ & $2.60\pm0.10$ \\
\end{tabular}
\end{center}
\end{table}
One may thus consider the tail exponents $\eta_-$ and $\eta_+$ to be
``hyper-universal'' as the actual values depend mostly on the ensemble
and, within the numerical accuracy, almost not at all on the
dimensionality. The actual values of the tail exponents and the 
skewnesses are reported in Table~\ref{tab}.  However the actual
scaling functions of $P(E)$ are not quite the same, as is demonstrated
by Fig.~\ref{fig3}(b) where the skewnesses of the scaled $P(E)$'s are
shown for ``single valley'' ensembles, and compared to the skewness
for various $N_z$ or $L_z$. As expected from the Gaussian-to-Gumbel
cross-over the skewnesses vary continuously. We finish with
Fig.~\ref{fig3}(c), which shows reasonably conclusively in
(2+1)-dimensions, the average roughness, another critical property, is
little if at all affected by the ensemble. What the figure
demonstrates is the insensitivity of the amplitude $a(N_z)$ in $w =
a(N_z) L^\zeta$, $\zeta=0.41$ to $N_z$ once $N_z > 1$. This is
in contrast to the average energy \cite{SAD01}.

To conclude we have here studied the statistics of the ground state
energy of critical elastic manifolds in random systems. These are the
first numerical calculations of the energy distribution, when $D>1$.
We show that the scaling function of the energy is a continuous
function of the external control parameter, the system ``thickness''
$N_z$. Thus the properties of the tails of $P(E)$ depend crucially on
such details, but not so much on dimensionality or in other words the
actual scaling exponents of the problem. One should note that the
expected variation of $\eta_-$, $\eta_+$ is typically logarithmic in
$N_z$ using usual convergence rates of penultimate distribution
extreme statistics. Therefore, such exponents have rather ``arbitrary''
values.

There are three open problems that we can outline. First, is it indeed
mostly a mathematical accident that {\it e.g.} $\eta_- \sim 1.5 \dots
1.7$ for all $D$ and $n$ in the single-valley ensembles? The existence
of such tails tells that all critical manifolds might share some
general properties with the KPZ-equation related systems that can be
treated on an analytical basis. In particular the free energy or large
deviation functional could be in such a case non-local.  Second, it
would be interesting (but numerically extremely hard) to look at $T>0$
the free energy distribution in the general case, except say for the
$(1+1)$ dimensional directed polymers, which is relatively easy and has been
studied in the literature~\cite{KBM}.  Third, one particular example
of a ``critical manifold'' is given by random field Ising magnets
(RFIM) at the bulk phase transition, since here one can map the
problem to RB domain walls in $(D+1)$ dimensions. Our results imply
that the distribution of the ground state energy should prove to be an
interesting quantity, in three (3D) and four dimensions. This is so in
particular since there is in the much-debated 3D case substantial
numerical evidence that the aspect ratio of the system will affect the
domain structure (and thus perhaps the energy distribution).

\acknowledgments

This work has been supported by the Academy of Finland's Centre of
Excellence Programme. It was also performed under the auspices of the
U.S.\ Dept.\ of Energy at the University of California/Lawrence
Livermore National Laboratory under contract no. W-7405-Eng-48 (ETS).
MJA would like to thank professors S. Coles and L. de Haan for 
correspondence on extremal statistics.

\end{document}